\documentclass[11pt]{article}

\pdfoutput=1

\usepackage[margin=1in]{geometry}
\usepackage[UKenglish]{babel}
\usepackage[T1]{fontenc}
\usepackage[utf8]{inputenc}
\usepackage{amsmath,amssymb,amsthm}
\usepackage{booktabs}
\usepackage{array}
\usepackage{graphicx}
\usepackage{url}
\usepackage[colorlinks=true,linkcolor=blue,citecolor=blue,urlcolor=blue]{hyperref}

\title{Calibrated Persistent Homology Tests for\\
High-dimensional Collapse Detection}

\author{Alexander Kalinowski\\
SUNY Empire University, USA\\
\href{mailto:alexander.kalinowski@sunyempire.edu}{alexander.kalinowski@sunyempire.edu}\\
\url{https://github.com/akalino}\\
\url{https://orcid.org/0000-0003-4961-628X}}

\date{}

\begin{document}

\maketitle

\begin{abstract}
We study detection of collapse in high-dimensional point clouds, where mass concentrates near a lower-dimensional set relative to a non-collapsed geometry. 
We propose persistent-homology-based test statistics under two well-studied filtrations, with cutoffs calibrated under a broad set of non-collapsed reference models.
We benchmark power across three alternative collapse mechanisms (linear/spectral, nonlinear-support, and contamination/heterogeneity) and distill the results into a mechanism map guiding the choice of filtration and statistic.
\end{abstract}


\noindent\textbf{Supplementary material:} \url{https://github.com/akalino/PH_Collapse_Detection}

\section{Motivation and Problem Proposal}
\label{sec:motivation}

High-dimensional collapse refers to a failure mode where a point cloud that is expected to be approximately isotropic becomes highly anisotropic, with mass concentrating near a lower-dimensional structure.
When point clouds lose intrinsic dimensional structure, downstream geometric reasoning can fail.
This phenomenon arises in neural network representation spaces, where collapse can degrade downstream performance.
Persistent homology (PH) is particularly useful when collapse is toward a curved or nonlinear support rather than a flat subspace.
We study detection of collapse from a single observed point cloud
$X=\{x_i\}_{i=1}^n\subset\mathbb{R}^d$ via statistical testing, seeking tests that detect collapse while controlling false positives under challenging non-collapsed null models.
We consider null families $\mathcal{N}$ intended to represent ``healthy'' geometry and alternatives $\mathcal{A}$ representing collapsed geometry with varying strength $\varepsilon$.
Our goal is to produce calibrated tests based on PH metrics that remain valid across a broad suite of null classes and that provide mechanism-level insight into when PH is informative.

\section{Contributions}
We develop and empirically evaluate PH-based hypothesis tests for collapse.
First, we construct calibrated tests using PH summaries under multiple null families and apply standard statistical corrections to control per-experiment error.
Second, we study two complementary scalar summaries, total persistence (TP) and mean tail excess (MTE), under both Vietoris–Rips (VR) and distance-to-a-measure (DTM) filtrations, and use the resulting mechanism map to highlight when each choice is informative.
Lastly, we organize alternatives into three collapse mechanisms: (A) linear/spectral collapse, (B) nonlinear-support collapse (concentration near curved structure), and (C) contamination/heterogeneity (mixtures and outliers).
A central output is a mechanism map in Table~\ref{tab:mechanism_map} summarizing which PH choices (TP vs.\ MTE; VR vs.\ DTM) are most effective by mechanism. 

Our work follows statistical TDA for stable persistence summaries and inference on persistence objects~\cite{cohensteiner2007stability,edelsbrunnerharer2010,carlsson2009topology}.
For uncertainty quantification and testing with persistence-based objects, we draw inspiration from confidence-set constructions and hypothesis testing for persistence diagrams~\cite{fasy2014confidence,robinsonturner2017ht}.
To improve robustness to contamination and heterogeneous sampling, we consider DTM-style constructions and robust topological inference tools \cite{chazal2011geometric,chazal2018robust}.

\section{Methodology: PH-Based Test Statistics and Calibration}
For $q\in\{0,1,2\}$, let $\mathrm{Dgm}_q(X)$ denote the $q$-dimensional persistence diagram computed from a filtration $K_\alpha(X)$.
We use two filtrations: (i) the Vietoris--Rips filtration and (ii) a DTM-based filtration, intended to be more robust to contamination and heterogeneous sampling.
From $\mathrm{Dgm}_q(X)$ we form two scalar summaries.
Let $\ell=d-b$ denote lifetime for a point $(b,d)\in\mathrm{Dgm}_q(X)$ and let $L_q(X)$ be the lifetimes in $\mathrm{Dgm}_q(X)$.



We define the total persistence (TP) and the mean tail excess (MTE).
For $p\ge 1$ and $\tau>0$, define
\[
\mathrm{TP}_{q,p}(X)=\sum_{\ell\in L_q(X)} \ell^p,
\qquad
\mathrm{MTE}_{q,\tau}(X)=
\frac{1}{|\{\ell\in L_q(X):\ell>\tau\}|}
\sum_{\ell\in L_q(X):\ell>\tau}(\ell-\tau),
\]
with $\mathrm{MTE}_{q,\tau}(X)=0$ if no lifetime exceeds $\tau$.

We fix a finite collection of summaries (choices of $q$, filtration, and statistic) and compute the corresponding values on $X$.
For each null family $\mathcal{N}$ and each summary $T$, we calibrate a cutoff by drawing $X^{(1)},\dots,X^{(B)}\sim\mathcal{N}$ and 
select the empirical $(1 - \alpha)$-quantile as the cutoff value.
We calibrate upper-tail cutoffs and reject for large values of $T(X)$. 
We benchmark power over collapse strength $\varepsilon$ across our three mechanisms and summarize outcomes in a mechanism map (Table~\ref{tab:mechanism_map}).

\section{Preliminary results}
We focus on our mechanism map; additional details on calibration and power are left to Appendix~\ref{sec:appendix}.
Table~\ref{tab:mechanism_map} summarizes power by mechanism, where MTE is the most consistently sensitive metric (especially for nonlinear-support collapse) and DTM generally improves robustness relative to VR.
TP contributes significantly less than MTE for both VR and DTM, indicating that a global metric of all persistence tracks too many unchanging features that may not accurately reflect collapse behavior.
This limitation could also be due to point cloud size; our initial experiments on clouds of dimension $d \in \{5, 10, 20\}$ and $n \in \{10, 50, 100\}$, motivating expansion of this grid for improving robustness and expanding the mechanism map.
Additionally, we will consider tracking intermediate steps as the healthy geometries migrate toward a collapsed state, matching the motivation of detecting poor training in neural network models.

\begin{table}[th!]
\centering
\caption{Mechanism map summary. Mean power (empirical rejection rate) is averaged over the datasets within each mechanism, over the (n, d, $\varepsilon$) grid, and by test (filtration × statistic). The bold values report the best performing test per mechanism.}
\label{tab:mechanism_map}
\setlength{\tabcolsep}{6pt}
\renewcommand{\arraystretch}{1.15}
\begin{tabular}{lcccc}
\textbf{Mechanism} & \textbf{VR-TP} & \textbf{VR-MTE} & \textbf{DTM-TP} & \textbf{DTM-MTE} \\
A & 0.006 & 0.229 & 0.010 & \textbf{0.257} \\
B & 0.010 & 0.525 & 0.048 & \textbf{0.548} \\
C & 0.018 & 0.173 & 0.016 & \textbf{0.192} \\
\end{tabular}
\end{table}



\newpage

\bibliographystyle{plain}
\bibliography{calibrated_ph}

@book{edelsbrunnerharer2010,
  title        = {Computational Topology: An Introduction},
  author       = {Edelsbrunner, Herbert and Harer, John},
  year         = {2010},
  publisher    = {American Mathematical Society},
  series       = {Mathematical Surveys and Monographs},
  volume       = {69},
  isbn         = {9780821849255}
}

@article{cohensteiner2007stability,
  title        = {Stability of Persistence Diagrams},
  author       = {Cohen-Steiner, David and Edelsbrunner, Herbert and Harer, John},
  journal      = {Discrete \& Computational Geometry},
  volume       = {37},
  number       = {1},
  pages        = {103--120},
  year         = {2007},
  doi          = {10.1007/s00454-006-1276-5}
}

@article{carlsson2009topology,
  title        = {Topology and Data},
  author       = {Carlsson, Gunnar},
  journal      = {Bulletin of the American Mathematical Society},
  volume       = {46},
  number       = {2},
  pages        = {255--308},
  year         = {2009},
  doi          = {10.1090/S0273-0979-09-01249-X}
}

@article{chazal2011geometric,
  title        = {Geometric Inference for Probability Measures},
  author       = {Chazal, Fr{\'e}d{\'e}ric and Cohen-Steiner, David and M{\'e}rigot, Quentin},
  journal      = {Foundations of Computational Mathematics},
  volume       = {11},
  number       = {6},
  pages        = {733--751},
  year         = {2011},
  doi          = {10.1007/s10208-011-9098-0}
}

@article{fasy2014confidence,
  title        = {Confidence Sets for Persistence Diagrams},
  author       = {Fasy, Brittany Terese and Lecci, Fabrizio and Rinaldo, Alessandro and Wasserman, Larry
                  and Balakrishnan, Sivaraman and Singh, Aarti},
  journal      = {The Annals of Statistics},
  volume       = {42},
  number       = {6},
  pages        = {2301--2339},
  year         = {2014},
  doi          = {10.1214/14-AOS1252}
}

@article{robinsonturner2017ht,
  title        = {Hypothesis Testing for Topological Data Analysis},
  author       = {Robinson, Andrew and Turner, Katharine},
  journal      = {Journal of Applied and Computational Topology},
  volume       = {1},
  number       = {2},
  pages        = {241--261},
  year         = {2017},
  doi          = {10.1007/s41468-017-0008-7}
}

@article{chazal2018robust,
  title        = {Robust Topological Inference: Distance To a Measure and Kernel Distance},
  author       = {Chazal, Fr{\'e}d{\'e}ric and Fasy, Brittany and Lecci, Fabrizio and Michel, Bertrand
                  and Rinaldo, Alessandro and Wasserman, Larry},
  journal      = {Journal of Machine Learning Research},
  volume       = {18},
  number       = {159},
  pages        = {1--40},
  year         = {2018},
  url          = {https://jmlr.org/papers/v18/15-484.html}
}

\newpage

\appendix

\section{Preliminary results}\label{sec:appendix}

\subsection{Reproducibility}

We provide our codebase at \url{https://github.com/akalino/PH_Collapse_Detection} for reproducibility purposes.
All experiments were run using 12GB of RAM and parallel processing across four cores.
The entire experiment can be run using the provided bash script; we additionally include the calibration file $tau\_map.csv$ as calibration is the most compute and time intensive step.

\subsection{Calibration table}

Table~\ref{tab:calibration} shows that calibration yields near-nominal error rates over a broad null suite.
The hardest nulls in our current grid are high-noise, manifold-like, and moderately anisotropic families (e.g., noisy sphere with $\sigma=0.5$ and elliptical Gaussian with $\eta \in \{0.1, 0.2\}$).

\begin{table}[th!]
\centering
\caption{Null calibration. Entries are mean empirical rejection rates averaged over the (n,d) grid for each null class and each test (filtration × statistic).}
\label{tab:calibration}
\setlength{\tabcolsep}{6pt}
\renewcommand{\arraystretch}{1.15}
\begin{tabular}{lcccc}
\toprule
\textbf{Point cloud} & \textbf{VR-TP} & \textbf{VR-MTE} & \textbf{DTM-TP} & \textbf{DTM-MTE} \\
\midrule
Standard Gaussian & 0.005 & 0.030 & 0.011 & 0.023 \\
Gaussian ($\eta = 0.05$) & 0.007 & 0.028 & 0.012 & 0.022 \\
Gaussian ($\eta = 0.1$) & 0.008 & 0.030 & 0.013 & 0.026 \\
Gaussian ($\eta = 0.2$) & 0.006 & 0.031 & 0.014 & 0.027 \\
Gaussian ($\eta = 0.5$) & 0.007 & 0.027 & 0.012 & 0.023 \\
Gaussian ($\eta = 1.0$) & 0.008 & 0.027 & 0.011 & 0.025 \\
Noisy sphere ($\sigma = 0.1$) & 0.050 & 0.000 & 0.018 & 0.000 \\
Noisy sphere ($\sigma = 0.3$) & 0.033 & 0.000 & 0.033 & 0.007 \\
Noisy sphere ($\sigma = 0.5$) & 0.038 & 0.000 & 0.042 & 0.019 \\
\bottomrule
\end{tabular}
\end{table}

\subsection{Power}

Table~\ref{tab:power_primary} reports primary-test power versus $\varepsilon$, indicating that detectable collapse in our current grid concentrates at larger values, motivating larger experiments on $(n, d, \varepsilon)$.

\begin{table}[th!]
\centering
\caption{Primary-test power versus collapse strength $\varepsilon$. Entries are mean power averaged over the (n,d) grid.}
\label{tab:power_primary}
\setlength{\tabcolsep}{6pt}
\renewcommand{\arraystretch}{1.15}
\begin{tabular}{lcccccccc}
\toprule
\textbf{Point cloud} & \textbf{Mechanism}& \textbf{0.05} & \textbf{0.1} & \textbf{0.2} & \textbf{0.5} & \textbf{1.0} & \textbf{1.5} & \textbf{2.0} \\
\midrule
$k$-plane & A & 0.000 & 0.000 & 0.000 & 0.000 & 0.008 & 0.013 & 0.030 \\
Spiked Gaussian & A & 0.002 & 0.000 & 0.000 & 0.000 & 0.002 & 0.003 & 0.020 \\
Swiss roll & B & 0.012 & 0.005 & 0.012 & 0.015 & 0.010 & 0.025 & 0.027 \\
Torus & B & 0.000 & 0.000 & 0.000 & 0.000 & 0.003 & 0.012 & 0.050 \\
Paraboloid & B & 0.000 & 0.000 & 0.000 & 0.000 & 0.012 & 0.020 & 0.018 \\
Contaminated $k$-cube & C & 0.000 & 0.000 & 0.000 & 0.000 & 0.012 & 0.015 & 0.048 \\
Contaminated $k$-plane & C & 0.000 & 0.000 & 0.000 & 0.000 & 0.017 & 0.003 & 0.012 \\
Contaminated sphere & C & 0.052 & 0.037 & 0.030 & 0.037 & 0.038 & 0.045 & 0.032 \\

\bottomrule
\end{tabular}
\end{table}

\end{document}